\documentclass[conference]{IEEEtran}
\IEEEoverridecommandlockouts

\usepackage{cite}
\usepackage{amsmath,amssymb,amsfonts}

\usepackage{algorithm}
\usepackage[noend]{algpseudocode}
\usepackage{listings}
\usepackage{graphicx}
\usepackage{textcomp}
\usepackage{xcolor}
\usepackage{hyperref} %
\usepackage{subcaption}
\usepackage{enumitem} %
\usepackage{array,multirow,graphicx}

\usepackage{slashbox}

\def\BibTeX{{\rm B\kern-.05em{\sc i\kern-.025em b}\kern-.08em
 T\kern-.1667em\lower.7ex\hbox{E}\kern-.125emX}}

\newcommand{\rottext}[1]{\rotatebox[origin=c]{90}{#1}}
\begin{document}

\lstset{language=Python,
 basicstyle=\ttfamily\scriptsize,
 keywordstyle=\bfseries,
 showstringspaces=false,
 morekeywords={net, convolutional, batch_normalize}
}

\title{Orpheus: A New Deep Learning Framework for Easy Deployment and Evaluation of Edge Inference}

\author{Perry Gibson, Jos\'e Cano \\
\emph{School of Computing Science, University of Glasgow, UK}
}

\maketitle

\begin{abstract}

Optimising deep learning inference across edge devices and optimisation targets such as inference time, memory footprint and power consumption is a key challenge due to the ubiquity of neural networks. Today, production deep learning frameworks provide useful abstractions to aid machine learning engineers and systems researchers. However, in exchange they can suffer from compatibility challenges (especially on constrained platforms), inaccessible code complexity, or design choices that otherwise limit research from a systems perspective. This paper presents Orpheus, a new deep learning framework for easy prototyping, deployment and evaluation of inference optimisations. Orpheus features a small codebase, minimal dependencies, and a simple process for integrating other third party systems. We present some preliminary evaluation results.

\end{abstract}

\section{Introduction}

Deploying deep learning applications like image classification on edge devices (e.g. IoT boards, smartphones, drones) is a challenging task, since large Deep Neural Networks (DNNs) do not fit on these devices or have unacceptable performance. Turner et al.~\cite{iiswc_2018} showed that machine learning optimisations like model compression may not work as expected at system level where one of the main metrics considered is the inference time. However, implementing and evaluating machine learning techniques on well-known frameworks like PyTorch or TensorFlow while targeting edge devices can be very tedious and challenging, as most frameworks are complex pieces of software with many requirements and dependencies. %

\begin{figure}[t]
 \centering
 \includegraphics[width=0.84\columnwidth]{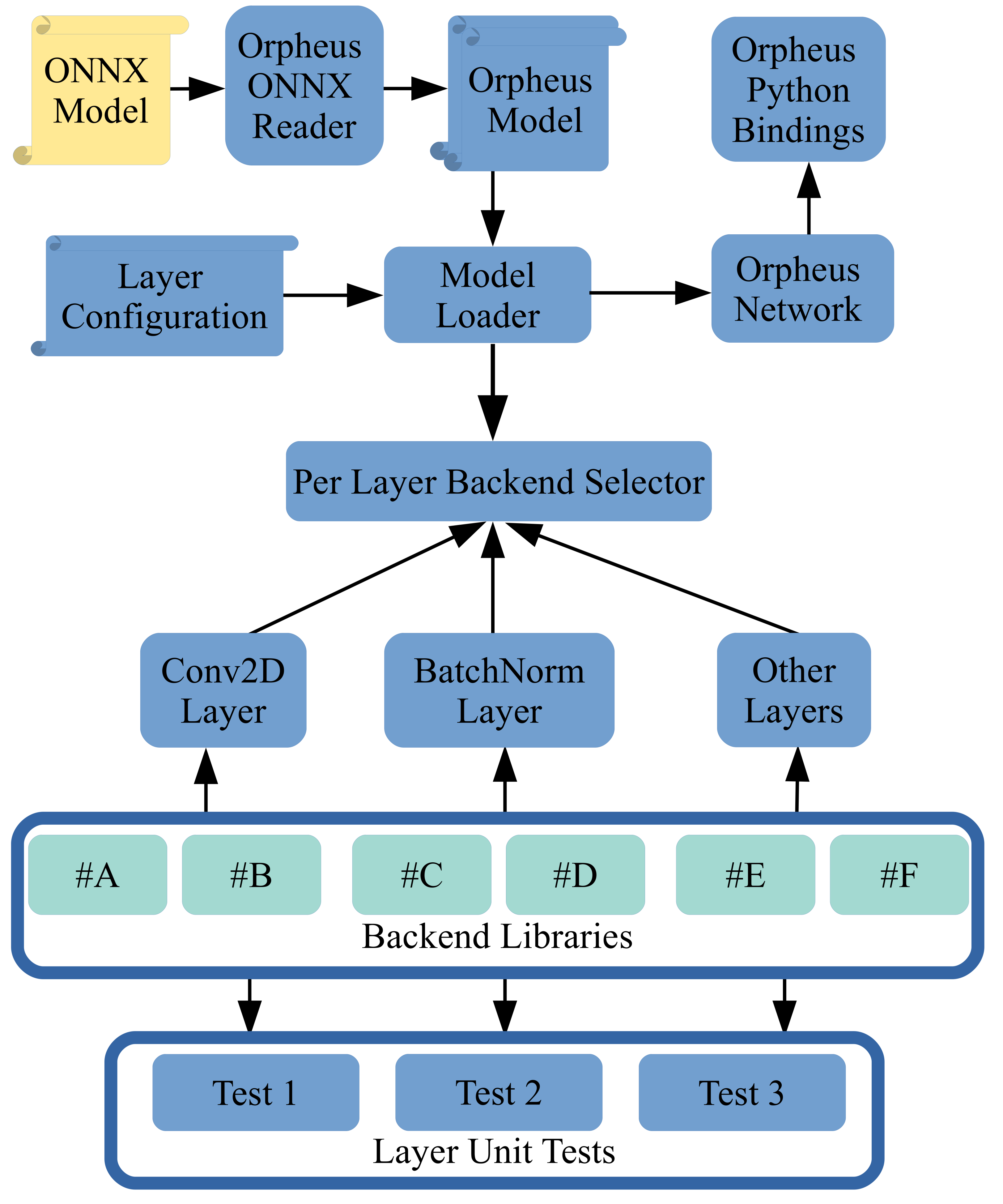}
 \caption{\label{fig:overview} Overview of Orpheus design. }
\end{figure}

In this paper we present Orpheus, a new research environment for exploring optimisations of neural network inference. The Orpheus framework is designed to support straightforward integration of different backends such as OpenCL kernels or third party libraries such as ARM Compute Library, and features infrastructure to load and evaluate models exported from other training frameworks. Therefore, we can obtain inference times for different backends and multiple models quickly, from a single programming interface. Our work is focused solely on inference, though the programming model does not preclude training. The main parts of Orpheus and contributions of this paper are as follows (Figure \ref{fig:overview}):

\begin{itemize}[noitemsep, topsep=0pt]

\item A simple and extensible programming model for comparing multiple neural network layer implementations in a consistent environment.

\item A system to parse pre-trained models exported to the ONNX format from popular training frameworks, and to apply simplifications to the computation graph.

\item Custom implementations of common neural network operations in C++ with alternative algorithms that can leverage APIs such as OpenMP. %

\item Easy integration of third party backends like Intel DNNL or Arm Compute Library.

\item Suite of unit tests to ensure correctness of all operations, and to provide ready-made assistance in the development and integration of new backends.

\item Infrastructure to run multiple inference experiments, evaluating full networks, and individual layers with the option of using Python bindings.
\end{itemize}

\section{Comparison of Deep Learning Frameworks}
\label{sec:dnn_frameworks}

Table \ref{table:frameworks} compares different deep learning frameworks according to key features that a platform for systems research should provide (we rate features 1-3 based on our experience):

\begin{itemize}
 
 \item Low-level modifications: Ability to easily access lower-level features of the platform (e.g. SIMD intrinsics).
 
 \item Model interoperability: Support for models trained in other frameworks, usually via the ONNX format.
 
 \item  Platform compatibility: Ease of deployment on a range of edge devices (e.g. with different hardware features).

 \item Codebase accessibility: Ability for users to prototype, test, and integrate new features or backends.
 
 \item Performance (inference time): Execution time to evaluate data on common neural networks, e.g. classify images.
 
\end{itemize}

\begin{table}[]
\centering
\caption{Comparison of Deep Learning frameworks.}
%\begin{tabular}{| p{35mm} || l | l | l | l | l |}
\begin{tabular}{| l || l | l | l | l | l |}
\hline
    \backslashbox[35mm]{\textbf{Feature}}{\textbf{Framework}} & \rottext{TF-Lite} & \rottext{PyTorch} & \rottext{DarkNet} & \rottext{TVM} & \rottext{Orpheus} \\ \hline
    Low-level modifications & 1 & 1 & 2 & 2 & 3 \\ \hline
    Model interoperability & 2 & 3 & 1 & 3 & 3 \\ \hline
    Platform Compatibility & 3 & 2 & 3 & 3 & 3 \\ \hline
    Codebase accessibility & 1 & 2 & 3 & 1 & 3 \\ \hline
    Performance (inference time) & 2 & 2 & 1 & 2 & 3 \\ \hline
\end{tabular}
\label{table:frameworks}
\end{table}

\textbf{TensorFlow-Lite}~\cite{tflite} (TF-Lite) is a version of TensorFlow's engine with a reduced set of operations, for mobile and IoT devices. We found it difficult to work with, due to its lack of clear documentation and limited operator support. Importing models is an error prone process, and some models (e.g. ResNets) have operations which are not supported. It can be accessed via a Python API, or integrated as a library.

%It is accessed via a Python API, or can be integrated into larger applications.

\textbf{PyTorch}~\cite{paszke2017automatic} supports model design, training, and inference via a high-level Python API. It is ideal for prototyping network architectures, and deploying them to server-class machines. However, the high-level API creates barriers to exploring performance issues, and making low-level modifications.

\textbf{DarkNet}~\cite{darknet13} is written in C and CUDA. Its small codebase makes it more accessible to change than other frameworks, and has minimal dependencies. However, it lacks competitive performance, and cannot import third party models.

\textbf{TVM}~\cite{tvm} is an open deep learning compiler stack. It provides competitive performance across a variety of platforms on some benchmarks. However, to leverage it fully, developers must become familiar with its niche programming model, and we have found areas where it performs poorly (e.g. replacing standard convolutional blocks with cheaper ones~\cite{moonshine}).

\textbf{Orpheus} is an inference-only framework written in C++. Its main design goal is to transparently support experimentation with alternative backends. In Orpheus, layers are treated as first class citizens, and have multiple implementations which are selected at runtime.

\section{Evaluation}
\label{sec:exp}

We now present some initial experimental results that compare Orpheus against some of the deep learning frameworks described in Section \ref{sec:dnn_frameworks}. We analyse the inference time of five DNN models (WRN-40-2, MobileNetV1, RestNet-18, Inception-v3, ResNet-50) on the CPU of the HiKey 970 board (Arm Cortex-A73). Note that we consider only one core.

\begin{figure}[t]
 \centering
   \includegraphics[width=0.95\columnwidth]{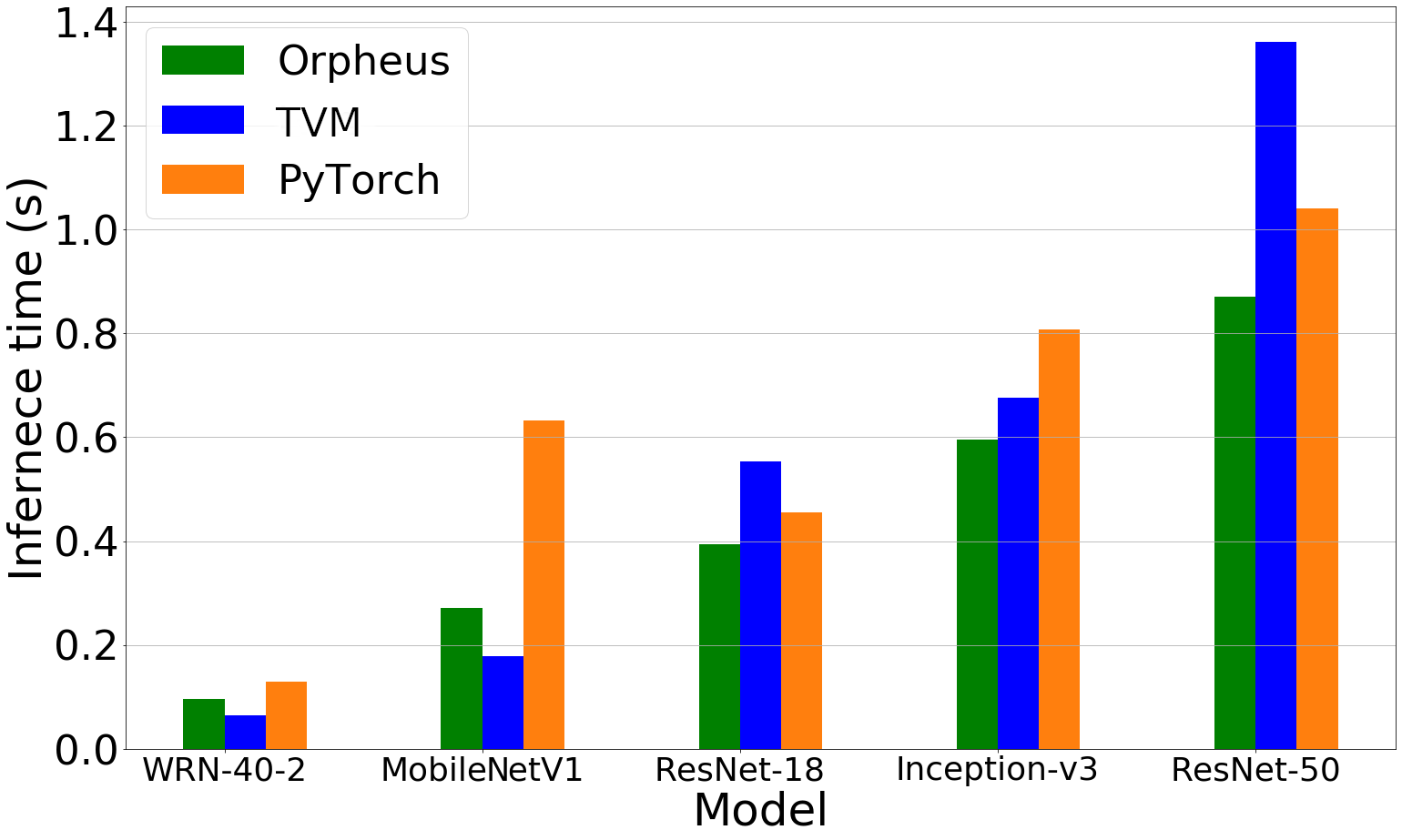}
 \caption{Inference time (1 thread) for the five network models. }
 \label{fig:cnn_inference}

\end{figure}

As we can see in Figure \ref{fig:cnn_inference}, Orpheus provides the best results for the biggest models (ResNets and Inception), whereas TVM is the best for the smallest ones (WRN and MobileNet). These results make sense, as Orpheus uses GEMM (General Matrix Multiply) convolution, which pays off for the larger matrices of the big models, while TVM uses ``spatial pack'' convolution which seems to work better for the small ones. PyTorch also uses GEMM, although its times are worse than Orpheus. 

We also see that PyTorch performs poorly for MobileNetV1 because of an inefficient implementation of the depthwise convolution. Finally, note that we do not include results for DarkNet and TF-Lite. For DarkNet, only the ResNet models were available and had inference time measured in seconds (e.g. $\sim$3s for ResNet-18). For TF-Lite, all models excepting ResNets were available but the Python API always selects the maximum number of threads, so we could not select one.

\section{Conclusion}

This paper presented Orpheus, a new deep learning framework that can enable future systems research for DNNs, due to its flexible design. Orpheus has already demonstrated value in our research, where standard frameworks posed challenges for lower level systems investigation. Orpheus provides a way for researchers to export trained neural networks to a transparent inference runtime, where components can be independently altered and assayed to answer various research questions. Additionally, Python bindings improve the ease of use for embedding in other experimental workflows.

\section*{Acknowledgment}

This work was supported by the European Union's Horizon 2020 research and innovation programme under grant agreement No 732204 (Bonseyes), and by the Swiss State Secretariat for Education, Research and Innovation (SERI) under contract number 16.0159.

\bibliography{ispass2020}

\begin{thebibliography}{1}
\providecommand{\url}[1]{#1}
\csname url@samestyle\endcsname
\providecommand{\newblock}{\relax}
\providecommand{\bibinfo}[2]{#2}
\providecommand{\BIBentrySTDinterwordspacing}{\spaceskip=0pt\relax}
\providecommand{\BIBentryALTinterwordstretchfactor}{4}
\providecommand{\BIBentryALTinterwordspacing}{\spaceskip=\fontdimen2\font plus
\BIBentryALTinterwordstretchfactor\fontdimen3\font minus
  \fontdimen4\font\relax}
\providecommand{\BIBforeignlanguage}[2]{{%
\expandafter\ifx\csname l@#1\endcsname\relax
\typeout{** WARNING: IEEEtran.bst: No hyphenation pattern has been}%
\typeout{** loaded for the language `#1'. Using the pattern for}%
\typeout{** the default language instead.}%
\else
\language=\csname l@#1\endcsname
\fi
#2}}
\providecommand{\BIBdecl}{\relax}
\BIBdecl

\bibitem{iiswc_2018}
J.~Turner, J.~Cano, V.~Radu, E.~J. Crowley, M.~O’Boyle, and A.~Storkey,
  ``Characterising across-stack optimisations for deep convolutional neural
  networks,'' in \emph{2018 IEEE International Symposium on Workload
  Characterization (IISWC)}, September 2018.

\bibitem{tflite}
``\BIBforeignlanguage{en}{{{TensorFlow Lite}}},''
  \url{https://www.tensorflow.org/lite/}.

\bibitem{paszke2017automatic}
A.~Paszke, S.~Gross, S.~Chintala, G.~Chanan, E.~Yang, Z.~DeVito, Z.~Lin,
  A.~Desmaison, L.~Antiga, and A.~Lerer, ``Automatic differentiation in
  {PyTorch},'' in \emph{NIPS Autodiff Workshop}, 2017.

\bibitem{darknet13}
J.~Redmon, ``Darknet: Open source neural networks in c,''
  \url{http://pjreddie.com/darknet/}, 2013--2016.

\bibitem{tvm}
T.~Chen, T.~Moreau, Z.~Jiang, L.~Zheng, E.~Yan, M.~Cowan, H.~Shen, L.~Wang,
  Y.~Hu, L.~Ceze, C.~Guestrin, and A.~Krishnamurthy, ``Tvm: An automated
  end-to-end optimizing compiler for deep learning,'' in \emph{Proceedings of
  the 12th USENIX Conference on Operating Systems Design and Implementation},
  ser. OSDI'18, 2018, pp. 579--594.

\bibitem{moonshine}
E.~J. Crowley, G.~Gray, and A.~Storkey, ``Moonshine: Distilling with cheap
  convolutions,'' in \emph{NeurIPS}, 2018.

\end{thebibliography}
\bibliographystyle{IEEEtran}

\end{document}